\documentclass[prl,twocolumn,superscriptaddress,showpacs,amssymb,amsmath,amsfonts,aps]{revtex4} 
\usepackage{graphicx} 
\begin{document} 
\newcommand{\thp}{$\Theta^+$} 
\newcommand{\lamstar}{$\Lambda$(1520)} 
\title{Search for the \thp~pentaquark in the reaction  
	$\gamma d \to pK^-K^+n$} 
 
\newcommand*{\ASU}{Arizona State University, Tempe, Arizona 85287-1504} 
\affiliation{\ASU} 
\newcommand*{\UCLA}{University of California at Los Angeles, Los Angeles, California  90095-1547} 
\affiliation{\UCLA} 
\newcommand*{\CMU}{Carnegie Mellon University, Pittsburgh, Pennsylvania 15213} 
\affiliation{\CMU} 
\newcommand*{\CUA}{Catholic University of America, Washington, D.C. 20064} 
\affiliation{\CUA} 
\newcommand*{\SACLAY}{CEA-Saclay, Service de Physique Nucl\'eaire, F91191 Gif-sur-Yvette, France} 
\affiliation{\SACLAY} 
\newcommand*{\CNU}{Christopher Newport University, Newport News, Virginia 23606} 
\affiliation{\CNU} 
\newcommand*{\CSU}{California State University, Dominguez Hills, Carson, CA 90747 } 
\affiliation{\CSU} 
\newcommand*{\UCONN}{University of Connecticut, Storrs, Connecticut 06269} 
\affiliation{\UCONN} 
\newcommand*{\ECOSSEE}{Edinburgh University, Edinburgh EH9 3JZ, United Kingdom} 
\affiliation{\ECOSSEE} 
\newcommand*{\FIU}{Florida International University, Miami, Florida 33199} 
\affiliation{\FIU} 
\newcommand*{\FSU}{Florida State University, Tallahassee, Florida 32306} 
\affiliation{\FSU} 
\newcommand*{\GWU}{The George Washington University, Washington, DC 20052} 
\affiliation{\GWU} 
\newcommand*{\ECOSSEG}{University of Glasgow, Glasgow G12 8QQ, United Kingdom} 
\affiliation{\ECOSSEG} 
\newcommand*{\ISU}{Idaho State University, Pocatello, Idaho 83209} 
\affiliation{\ISU} 
\newcommand*{\INFNFR}{INFN, Laboratori Nazionali di Frascati, 00044 Frascati, Italy} 
\affiliation{\INFNFR} 
\newcommand*{\INFNGE}{INFN, Sezione di Genova, 16146 Genova, Italy} 
\affiliation{\INFNGE} 
\newcommand*{\ORSAY}{Institut de Physique Nucleaire ORSAY, Orsay, France} 
\affiliation{\ORSAY} 
\newcommand*{\ITEP}{Institute of Theoretical and Experimental Physics, Moscow, 117259, Russia} 
\affiliation{\ITEP} 
\newcommand*{\JMU}{James Madison University, Harrisonburg, Virginia 22807} 
\affiliation{\JMU} 
\newcommand*{\UK}{University of Kentucky, Lexington, Kentucky 40506} 
\affiliation{\UK} 
\newcommand*{\KYUNGPOOK}{Kyungpook National University, Daegu 702-701, Republic of Korea} 
\affiliation{\KYUNGPOOK} 
\newcommand*{\UMASS}{University of Massachusetts, Amherst, Massachusetts  01003} 
\affiliation{\UMASS} 
\newcommand*{\MOSCOW}{Moscow State University, General Nuclear Physics Institute, 119899 Moscow, Russia} 
\affiliation{\MOSCOW} 
\newcommand*{\UNH}{University of New Hampshire, Durham, New Hampshire 03824-3568} 
\affiliation{\UNH} 
\newcommand*{\NSU}{Norfolk State University, Norfolk, Virginia 23504} 
\affiliation{\NSU} 
\newcommand*{\NCAT}{North Carolina A \& T State University, Greensburo, North Carolina 27411} 
\affiliation{\NCAT}
\newcommand*{\UNCW}{University of North Carolina, Wilmington, North Carolina 28403} 
\affiliation{\UNCW}
\newcommand*{\OHIOU}{Ohio University, Athens, Ohio  45701} 
\affiliation{\OHIOU} 
\newcommand*{\ODU}{Old Dominion University, Norfolk, Virginia 23529} 
\affiliation{\ODU} 
\newcommand*{\RPI}{Rensselaer Polytechnic Institute, Troy, New York 12180-3590} 
\affiliation{\RPI} 
\newcommand*{\RICE}{Rice University, Houston, Texas 77005-1892} 
\affiliation{\RICE} 
\newcommand*{\URICH}{University of Richmond, Richmond, Virginia 23173} 
\affiliation{\URICH} 
\newcommand*{\SCAROLINA}{University of South Carolina, Columbia, South Carolina 29208} 
\affiliation{\SCAROLINA} 
\newcommand*{\JLAB}{Thomas Jefferson National Accelerator Facility, Newport News, Virginia 23606} 
\affiliation{\JLAB} 
\newcommand*{\UNIONC}{Union College, Schenectady, NY 12308} 
\affiliation{\UNIONC} 
\newcommand*{\VT}{Virginia Polytechnic Institute and State University, Blacksburg, Virginia   24061-0435} 
\affiliation{\VT} 
\newcommand*{\VIRGINIA}{University of Virginia, Charlottesville, Virginia 22901} 
\affiliation{\VIRGINIA} 
\newcommand*{\WM}{College of William and Mary, Williamsburg, Virginia 23187-8795} 
\affiliation{\WM} 
\newcommand*{\YEREVAN}{Yerevan Physics Institute, 375036 Yerevan, Armenia} 
\affiliation{\YEREVAN} 
\newcommand*{\NOWUNH}{University of New Hampshire, Durham, New Hampshire 03824-3568} 
\newcommand*{\NOWCMU}{Carnegie Mellon University, Pittsburgh, Pennsylvania 15213} 
\newcommand*{\NOWSACLAY}{CEA-Saclay, Service de Physique Nucl\'eaire, F91191 Gif-sur-Yvette,Cedex, France} 
\newcommand*{\NOWUCLA}{University of California at Los Angeles, Los Angeles, California  90095-1547} 
\newcommand*{\NOWUMASS}{University of Massachusetts, Amherst, Massachusetts  01003} 
\newcommand*{\NOWMIT}{Massachusetts Institute of Technology, Cambridge, Massachusetts  02139-4307} 
\newcommand*{\NOWCUA}{Catholic University of America, Washington, D.C. 20064} 
\newcommand*{\NOWGEISSEN}{Physikalisches Institut der Universitaet Giessen, 35392 Giessen, Germany} 
 
\author {B.~McKinnon}  
\affiliation{\ECOSSEG} 
\author {K.~Hicks}  
\affiliation{\OHIOU} 
\author {N.A.~Baltzell}  
\affiliation{\SCAROLINA} 
\author {D.S.~Carman}  
\affiliation{\OHIOU} 
\author {M.D.~Mestayer}  
\affiliation{\JLAB} 
\author {T.~Mibe}  
\affiliation{\OHIOU} 
\author {M.~Mirazita}  
\affiliation{\INFNFR} 
\author {S.~Niccolai}  
\affiliation{\ORSAY} 
\author {P.~Rossi}  
\affiliation{\INFNFR} 
\author {S.~Stepanyan}  
\affiliation{\JLAB} 
% ...
\collaboration{The CLAS Collaboration} 
     \noaffiliation 
\date{\today} 
 
\begin{abstract} 
A search for the \thp~in the reaction $\gamma d \to pK^-K^+n$ 
was completed using the CLAS detector at Jefferson Lab.   
A study of the same reaction, published earlier, reported the 
observation of a narrow \thp~resonance.
The present experiment, with more than 30 times the integrated luminosity 
of our earlier measurement, does not show any evidence for a narrow 
pentaquark resonance.
The angle-integrated upper limit on \thp~production in the mass range of 
1.52 to 1.56 GeV/c$^2$ for the $\gamma d \to pK^-\Theta^+$ reaction is 
0.3 nb (95\% CL). This upper limit depends on assumptions made 
for the mass and angular distribution of \thp~production. 
Using \lamstar~production as an empirical measure of 
rescattering in the deuteron, the cross section upper limit for the 
elementary $\gamma n \to K^-\Theta^+$ reaction is estimated to be a 
factor of 10 higher, {\it i.e.}, $\sim 3$ nb (95\% CL).
\end{abstract} 
 
\pacs{12.39.Mk, 13.60.Rj, 14.20.Jn, 14.80.-j} 
 
\maketitle 
 
 
A tenet of the original quark model, first introduced by Gell-Mann 
and Ne'eman \cite{eightfold}, is that strongly-interacting 
particles have no more than three valence quarks.
Exotic multi-quark structures beyond the basic quark model were  
suggested by Jaffe \cite{jaffe77} and others in the 1970's, but  
it is now thought that these resonances are too wide to be detected 
by experiments. Recently, the prediction of a narrow pentaquark  
\cite{dpp97} and an initial report by the LEPS Collaboration 
\cite{leps03} have revitalized interest in searches for an  
exotic baryon with valence quark structure ($uudd\bar{s}$),  
known as the \thp.  If the \thp~exists, then 
this presents a challenge to theorists to find a way to 
describe pentaquarks, using either effective degrees of freedom 
\cite{jw03,ekp04} or lattice gauge calculations 
\cite{mathur04} based on QCD.
 
The reaction $\gamma d \to pK^-K^+n$ was measured by the 
CLAS Collaboration \cite{step03}, showing evidence for 
the \thp~decaying to $nK^+$ at a mass of about 1.54 GeV/c$^2$.
Considering the important implications of a possible pentaquark state, 
it was necessary to test the reproducibility of our previous result.
In addition, there are several experiments that see evidence for the 
\thp~and many that do not (see the reviews \cite{reviews}).
It is crucial to understand why some experiments see a peak and others 
do not \cite{karliner04}. For example, a strong peak identified as 
the \thp~was reported for $\gamma p \to \pi^+K^-K^+n$ in \cite{kubarov04}.
On the other hand, a high-statistics search for the \thp~in the 
$\gamma p \to K^0K^+n$ reaction reported a null result \cite{battag06}.
We present here a new search for the \thp~in the $\gamma d \to pK^-K^+n$ 
reaction with much improved statistical precision.


The present data were acquired during a two-month period in early 2004 
with the CLAS detector \cite{clas03} and the Hall B photon 
tagging system \cite{tagger}.  The incident electron beam energy was 
$E_0$=3.776~GeV, 
producing tagged photons in the range from 0.8 to 3.6~GeV.  
In order to ensure accuracy in the absolute mass scale 
for the current experiment,
the tagging spectrometer was calibrated independently of CLAS 
using the conversion of photons into $e^+e^-$ pairs in an 
aluminum foil.
The $e^+$ and $e^-$ were momentum-analyzed in a pair spectrometer 
having a precision field-mapped magnet, 
so that non-linear aspects of the photon energy measurement 
were calibrated, giving an accuracy of 0.1\%$\cdot$$E_\gamma$ 
over the full energy range.

The photon beam was directed onto a 24-cm long liquid-deuterium target.
One difference from our previous experiment \cite{step03} is that the 
target was placed 25 cm upstream of the center of the CLAS detector to 
increase the forward angle acceptance of negatively charged particles. 
The trigger required two charged particles detected in coincidence 
with a tagged photon.
The torus magnet was run at two settings, low field (2250 Amps) and 
high field (3375 Amps), each for about half of the run period.  
The low field setting has slightly better 
acceptance at forward angles, but worse momentum resolution.  The high 
field setting was the same as that used in Ref. \cite{step03}.  
The CLAS momentum resolution is on the order of 0.5-1.0\% (rms) depending 
on the kinematics.  Detailed calibrations of the CLAS detector subsystems 
were performed to achieve an accuracy of 1-2~MeV/c$^2$ in the 
$nK^+$ invariant mass distribution.  An integrated luminosity of about 
38~pb$^{-1}$ (for $E_\gamma > 1.5$ GeV) was collected here.

The quality of the detector calibrations and cross section 
normalization factors was verified using the reaction 
$\gamma d \to \pi^- pp$.
The differential cross sections in the photon energy range near 1.1 GeV 
were compared with the same reaction measured by the Hall A Collaboration 
\cite{zhu04}, and also the world data, 
with good agreement (within $\sim 10$\% over most of the angular range).

\begin{figure} 
\includegraphics[scale=0.5]{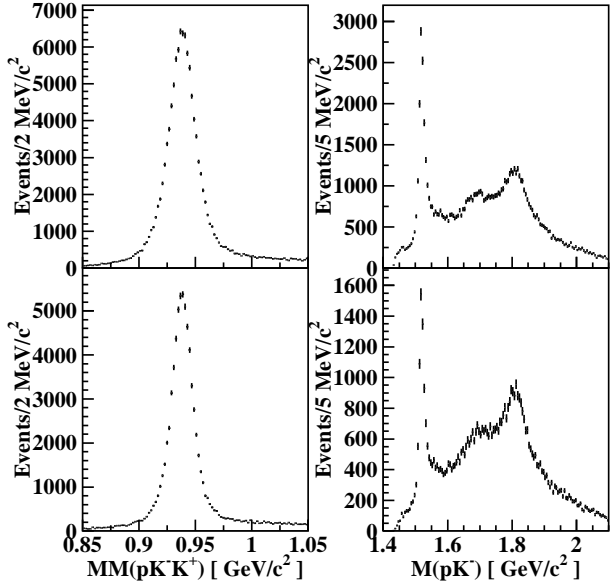} 
\caption{\label{fig:hi-lo} The raw data for the 
missing mass, $MM(pK^-K^+)$, of the reaction $\gamma d \to p K^- K^+ X$ 
showing a clean neutron peak (left). The 
invariant mass spectra of the detected  $pK^-$ (right) show the 
\lamstar~peak along with higher mass hyperons.
The data are shown separately for the low field (top) and high field  
(bottom) settings of the CLAS torus magnet. 
} 
\end{figure}

 
The event selection here is similar to that of Ref. \cite{step03}, 
requiring detection of 
one proton, one $K^+$, one $K^-$, and up to one neutral particle. 
All had vertex times within $\sim 1.0$ ns of the tagged photon. 
The missing mass of the $\gamma d \to p K^- K^+ X$ reaction and the 
invariant mass of the detected $pK^-$ particles, $M(pK^-)$, are shown 
in Fig. \ref{fig:hi-lo}.  The missing mass was required to be within 
$\pm 3\sigma$ of the neutron mass, where $\sigma$ is the mass resolution 
($\simeq 8$ MeV) of the peak shown.  Also, the missing momentum (not shown) 
was required to be greater than 0.20 GeV/c in order to remove spectator 
neutrons. Simulations of the decay $\Theta^+ \to nK^+$ show that this 
cut does not affect the \thp~detection efficiency.
Events corresponding to $\phi$-meson production were cut 
by requiring the $K^+K^-$ mass to be above 1.06 GeV/c$^2$, 
and similarly the \lamstar~was cut by removing events from 
$1.495<M(pK^-)<1.545$ GeV/c$^2$, see Fig. \ref{fig:hi-lo}.
Variations of these event selection cuts were studied and 
yield results consistent with those given below.

In Fig. \ref{fig:hi-lo}, both the low field setting (top) and the high 
field setting (bottom) are independent data sets.  The $M(pK^-)$ spectra  
show a prominent peak for the $\Lambda$(1520) and also strength at  
higher mass corresponding to well known $\Lambda^*$ resonances 
at  1.67, 1.69, and 1.82 GeV.  
These $\Lambda^*$ resonances are suppressed in the final data sample 
due to the neutron momentum cut, since in 
$\gamma p \to K^+ \Lambda^*$ the neutron is a spectator.
Also, $\Lambda^*$ production is not necessarily incompatible 
with \thp~production, as the $\gamma d \to \Lambda^* \Theta^+$
reaction still conserves strangeness. 
A narrow pentaquark peak with sufficient cross section would still be 
visible on top of the broad background from the $\Lambda^*$ resonances 
projected onto the $nK^+$ mass spectrum.  
 
\begin{figure} 
\includegraphics[scale=0.5]{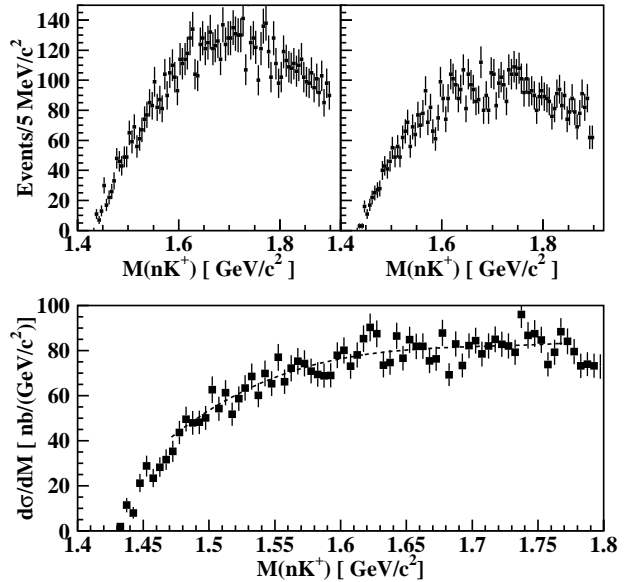} 
\caption{\label{fig:xsec} The invariant masses of the $nK^+$ system for 
both the low field (top left) and high field (top right) data sets, after 
applying all event selection cuts.  
The cross section per mass bin for the combined data (bottom) 
is shown along with a polynomial fit.
} 
\end{figure}

The spectra of the invariant mass of the $nK^+$ system, $M(nK^+)$, 
are shown in Fig. \ref{fig:xsec}, after applying the above analysis cuts. 
These spectra were constructed using the neutron mass as an explicit 
constraint (as contrasted with the missing mass of the $pK^-$ 
in Ref. \cite{step03} which did not use this constraint).  
A kinematic fitting approach gives nearly identical mass spectra
(but using a more elaborate procedure). 
The $M(nK^+)$ spectra in Fig. \ref{fig:xsec} do not show any 
evidence for a narrow peak near 1.54 GeV/c$^2$. 

The $M(nK^+)$ spectra at the top of Fig. \ref{fig:xsec} were then 
corrected for the CLAS detector acceptance and normalized by the 
luminosity, resulting in the combined data shown in the bottom plot.
The acceptance correction comes from a Monte Carlo simulation
that matches the exponential 
$t$-dependence of the measured $K^+$ and $K^-$ momenta, and
was fitted to the experimental neutron momentum distribution
in the range $p_n > 0.2$ GeV/c.  
Using this Monte Carlo, the angle-integrated acceptance of the CLAS 
detector ranges from 0.7\% to 1\% in the final data sample for both 
the high and low field settings.

The cross section spectrum of Fig. \ref{fig:xsec} (lower) was fit 
with a third-degree polynomial, as shown.  This curve  
was then held fixed and the excess (or deficit)  
above (or below) the polynomial was used to determine 
the cross section upper limit by standard methods \cite{feldman}.
The upper limit was also checked on both the low and high field data 
separately, giving consistent results.
We emphasize that this upper limit is for the $\gamma d \to pK^-\Theta^+$
reaction, and not for the elementary cross section on the neutron.
\begin{figure} 
\includegraphics[scale=0.5]{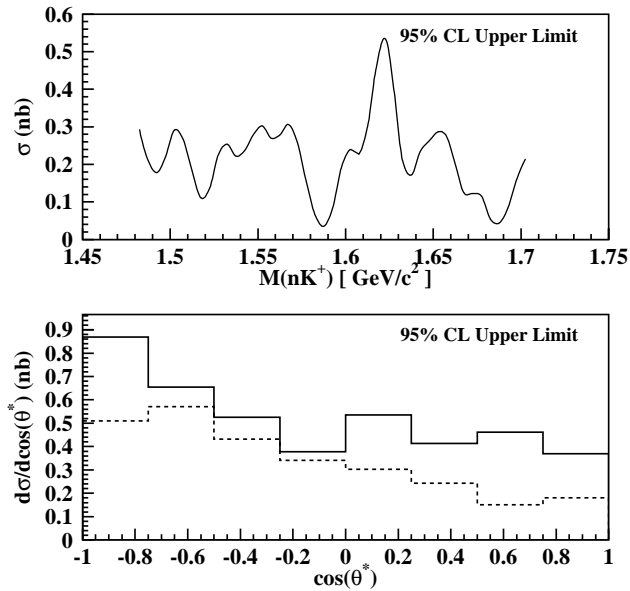} 
\caption{\label{fig:upper} The 95\% confidence limit (CL) 
upper limit cross sections for the $\gamma d \to pK^- \Theta^+$ reaction 
using the combined low and high field data.
The top plot is a function of mass, whereas 
the lower plot shows the maximum upper limit for the given $\theta^*$ bin,
for a mass range between 1.52 and 1.56 GeV/c$^2$ (solid) or just at mass 
1.54 GeV/c$^2$ (dashed).
}
\end{figure} 
 
The angle-integrated upper limit for the $\gamma d \to pK^-\Theta^+$ 
reaction, shown in Fig. \ref{fig:upper} (top),
is calculated to be less than 0.3 nb in the mass range of 1.5-1.6 GeV/c$^2$. 
This estimate assumes that our Monte Carlo corrects for the CLAS 
detector acceptance in unmeasured kinematic regions, 
such as for very forward angle kaons.  As mentioned above, our 
acceptance calculation assumes an exponential $t$-dependence, 
giving kaons preferentially at forward angles.

The most important dynamical variable that affects the acceptance is 
the angular dependence of the cross section.
The upper limit for the differential cross section is shown in the bottom 
half of Fig. \ref{fig:upper} as a function of $\cos\theta^*$, where 
$\theta^*$ is the angle of the $nK^+$ system, corresponding to the 
$\Theta^+$ decay particles, in the center-of-mass frame 
relative to the beam direction.  The 
solid line denotes the maximum upper limits in the mass range from 1.52 
to 1.56 GeV/c$^2$ for a given bin of $\cos\theta^*$.   
The dashed line is the upper limit at a particular mass of 1.54 GeV/c$^2$. 
The corresponding acceptance in CLAS covers the full angular region, 
and is only a factor of two smaller at forward angles of the $K^-$ 
($i.e.$ \thp~ at $\cos\theta^*<-0.75$) 
as compared with mid-range angles ($\cos\theta^*\sim 0$).

In determining the upper limit, systematic uncertainties need to 
be studied.  The dominant uncertainty is the 
unknown angular dependence of possible \thp~production. 
For example, if the \thp~were produced primarily with forward-angle 
$K^-$, as suggested by Ref. \cite{hosaka}, then we only need to 
consider the first angular bin in Fig. \ref{fig:upper} (lower). 
The upper limit (95\% CL) for the differential cross section, 
$d\sigma/d(\cos \theta^*)$, in this bin ($\cos \theta^*<-0.75$) 
is about 0.5 nb for a mass of 1.54 GeV/c$^2$, and after integrating 
over $\cos \theta^*$ (assuming zero contribution outside this bin), 
the upper limit on the \thp~total cross section is about 0.125 nb.  
The upper limit is well defined for a given set 
of assumptions about the \thp~mass, angular dependence, {\it etc.}, 
and we believe that the limits given above cover most of the 
reasonable alternatives.  Fluctuations in the upper limits 
due to other systematic effects are smaller \cite{hicks}.

The connection between our upper limits from 
deuterium and those from the elementary reaction on a free neutron 
is now explored.  There is no involvement of a proton   
in $\gamma n \to K^-\Theta^+ \to K^-K^+n$, 
but a proton was required in our deuterium analysis.  
To be detected in CLAS, a proton must have a momentum of at least 
0.35 GeV/c, and hence a mechanism to gain this momentum must be modeled.  
In order to set upper limits for \thp~production off the neutron, 
we need to correct for rescattering of the spectator proton in deuterium. 
For example, there could be  
rescattering of the $K^-$ from the proton.
 
\begin{figure} 
\includegraphics[scale=0.5]{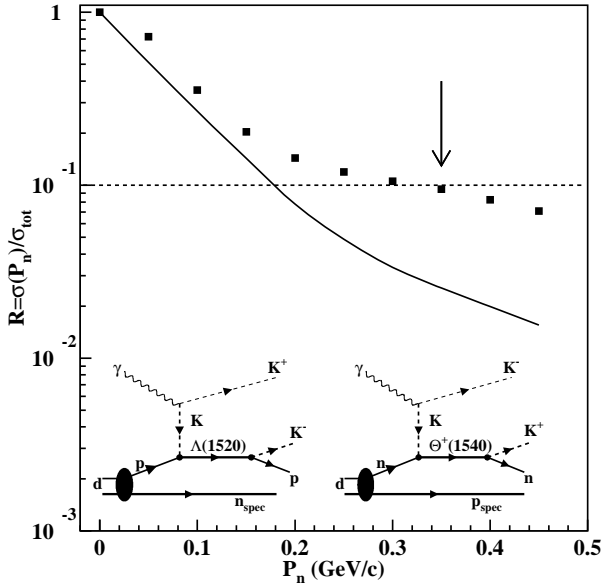} 
\caption{\label{fig:rescat} The relative cross section fraction 
for \lamstar~production  as a function of the neutron momentum
from our data (points).  
A cutoff of 350 MeV/c on the horizontal axis corresponds to a 
factor of 10 loss in detection of the \lamstar~compared with no cut. 
Also shown by the solid curve is the  
fraction calculated using only the Fermi motion of the neutron.  The  
reaction diagrams at the bottom show the symmetry between  
$t$-channel production for \lamstar~and \thp~reactions.
} 
\end{figure}

We use a phenomenological approach to estimate the 
rescattering of the spectator nucleon using the $t$-channel symmetry 
between $\Lambda$(1520) and \thp~production.
The first-order $t$-channel diagrams are shown at the bottom of 
Fig. \ref{fig:rescat}.  
For $\Lambda$(1520) production, the neutron is a spectator in 
this $t$-channel diagram, whereas for \thp~production, the 
proton is a spectator. 
Assuming that the $t$-channel dominates both $\Lambda$(1520) and 
\thp~production, a direct measure of rescattering of the neutron for 
$\Lambda$(1520) production can be used to estimate the 
amount of rescattering of the proton in \thp~production. 
This is a conservative estimate since the $K^+n$ 
scattering cross section is smaller than that for $K^-p$.
In addition, the \thp~would likely have a 
larger radius than the \lamstar, and thus a larger cross 
section for final state interactions. 

The fraction of cross section for $K^+ \Lambda(1520)$ production as a 
function of the neutron momentum cut is shown by the solid points in Fig. 
\ref{fig:rescat}. This ratio is unity when no momentum cut is applied, 
and only 10\% of the cross section survives for neutron momenta
greater than 0.35 GeV/c, as shown by the arrow.  
For comparison, the solid line in Fig. \ref{fig:rescat} shows the same 
fraction calculated using Fermi motion of the neutron in deuterium, 
based on the Paris potential \cite{paris}.  The points are higher than the 
line, indicating substantial rescattering due to final state interactions. 
Assuming a similar reduction factor in \thp~production  
for {\it protons} above 0.35 GeV/c, the upper limit  
for $\gamma n \to \Theta^+ K^-$ is estimated 
to be a factor of 10 higher than the upper limits presented 
in Fig. \ref{fig:upper}. 

A comparison of the current results with our previous report 
\cite{step03} is instructive. To do this, the 
current data were constrained, by software, to use the same 
event selection and the same photon energy region
as was used in Ref. \cite{step03}.  
Only the high field data are used here.  This analysis will 
be called the ``repeat study", since the analysis 
conditions are essentially unchanged from Ref. \cite{step03}.

\begin{figure} 
\includegraphics[scale=0.5]{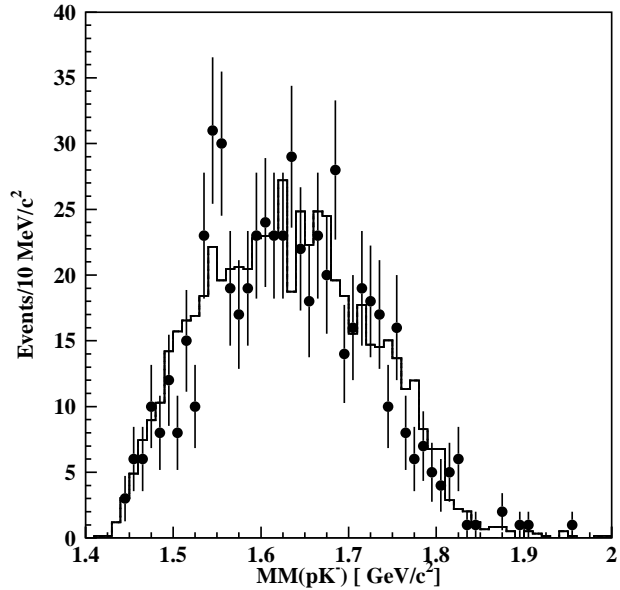} 
\caption{\label{fig:g2a} Comparison of the previously published 
\cite{step03} result (points) with the current result (histogram)  
normalized (by a factor of 1/5.92) to get the same total number of counts.} 
\end{figure}

In Fig. \ref{fig:g2a}, the results of Ref. \cite{step03} 
(points with statistical error bars) are compared with 
the results of the repeat study (histogram), rescaled 
for comparison by the ratio of the total counts.
The peak at 1.54 GeV/c$^2$ from Ref. \cite{step03} is not 
reproduced in the repeat study.
For comparison, the number of \lamstar~events scales, within 
statistical uncertainties, with the ratio of exclusive 
$pK^-K^+n$ events (current/previous).
\footnote{ The ratio is $6.6\pm0.2$ for $pK^-K^+n$ events 
and $6.3\pm 0.5$ for the \lamstar.}

Assuming that the histogram in Fig. \ref{fig:g2a} represents the 
true shape of the background, a smooth third-degree polynomial was 
fit to the histogram over the range from 1.45 to 1.75 GeV/c$^2$, 
giving a reduced $\chi^2 \simeq 1.15$.
A Gaussian peak, with width fixed at 7.0 MeV (the CLAS 
detector resolution) at a mass of 1.542 GeV/c$^2$, 
was fit on top of the polynomial 
background shape to the solid points in Fig. \ref{fig:g2a}, 
giving $25\pm 9$ counts in the peak.
The significance of a fluctuation is given by $S/\sqrt{B+V}$ 
where $S$ is the signal above background, $B$ is the background 
and $V$ is the variance in the background \cite{frodesen}.  
The variance of the background is difficult to estimate precisely, 
unless the shape is known.  Using the polynomial fit parameters, 
the values of $B$ and $V$ in the region from 
1.53-1.56 GeV/c$^2$ are 57 and 7, respectively.
This gives a new statistical significance of 3.1 $\sigma$.
In hindsight, the original signal size estimate of $5.2\pm 0.6\ \sigma$ 
in our previous publication was due to a significant 
under-estimate of the background.

 
In summary, 
the reaction $\gamma d \to pK^-K^+n$ has been measured using the 
CLAS detector where the neutron was identified by missing mass.  
A search for a narrow \thp~resonance decaying into $nK^+$ was done 
in the mass range of 1.48 to 1.7 GeV/c$^2$.    
The upper limit (95\% CL) for the total cross section of \thp~production 
ranges from 0.15-0.3 nb, depending on its angular distribution, 
for a mass of 1.54 GeV/c$^2$. 
An upper limit for the elementary process, $\gamma n \to K^-\Theta^+$, 
requires a correction for the proton momentum cutoff. 
The size of this correction 
was estimated from a phenomenological model based on
$\Lambda$(1520) production to be a factor of 10, giving an upper 
limit for the elementary $\gamma n \to K^-\Theta^+$ reaction 
estimated at $\sim 3$ nb.
The current null result shows that the $nK^+$ invariant mass 
peak of Ref. \cite{step03} could not be reproduced, and puts 
significant limits on the possible production cross section of a 
narrow \thp.
 
We would like to thank the staff of the Accelerator and Physics 
Divisions at Jefferson Lab who made this experiment possible. 
Acknowledgements for the support of this experiment go also to 
the Italian Istituto Nazionale de Fisica Nucleare, the French 
Centre National de la Recherche Scientifique and Commissariat \`a 
l'Energie Atomique, the Korea Research Foundation, 
the U.S. Department of Energy and the 
National Science Foundation, and the U.K. Engineering and 
Physical Science Research Council.  
The Southeastern Universities Research Association (SURA) operates the  
Thomas Jefferson National Accelerator Facility for the United States  
Department of Energy under contract DE-AC05-84ER40150.  
 
\bibliography{g10prl} 
 
\end{document}